# Effects of Static Magnetic Field on Growth of Leptospire, *Leptospira interrogans serovar canicola*: Immunoreactivity and Cell Division

WANNAPONG TRIAMPO,[1,4*] GALAYANEE DOUNGCHAWEE,[2] DARAPOND TRIAMPO,[3] JIRASAK WONG-EKKABUT,[1] and I-MING TANG[1,4]

*Department of Physics, Faculty of Science, Mahidol University, Bangkok 10400, Thailand,[1] Department of Pathobiology, Faculty of Science, Mahidol University, Bangkok 10400, Thailand,[2] Department of Chemistry, Faculty of Science, Mahidol University, Bangkok 10400, Thailand,[3] Capability Building Unit in Nanoscience and Nanotechnology, Faculty of Science, Mahidol University, Bangkok 10400, Thailand[4]*

*Corresponding author. e-mail: **scwtr@mahidol.ac.th** ; **wtriampo@yahoo.com**
phone:+662-889-2337   fax: +662-354-7159

**Abstract:** The effects of the exposure of the bacterium, *Leptospira interrogans* serovar *canicola* to a constant magnetic field with magnetic flux density from a permanent ferrite magnet = 140±5 mT were studied. Changes in *Leptospira* cells after their exposure to the field were determined on the basis of changes in their growth behavior and agglutination immunoreactivity with a homologous antiserum using dark-field microscopy together with visual imaging. The data showed that the exposed *Leptospira* cells have lower densities and lower agglutination immunoreactivity than the unexposed control group. Interestingly, some of the exposed *Leptospira* cells showed abnormal morphologies such as large lengths. We discussed some of the possible reasons for these observations.

# INTRODUCTION

Leptospirosis is an acute febrile illness caused by pathogenic spirochete bacteria of the genus *Leptospira* (1, 2). This disease has emerged as an important public health problem worldwide. The symptoms of this disease can range from mild-flu-like symptoms to severe (often fatal) complications such as renal and/or liver failure and hemorrhage (referred to as Weil's syndrome) (3). Most outbreaks tend to be seasonal in nature and are often associated with environmental factors, animals, and agricultural and occupational cycles such as rice cultivation in marshy lands. Mammals such as rats and cattle are commonly involved in the transmission of this disease to humans via direct or indirect exposure to contaminated tissues or urine (1, 2, 4). Out-breaks of leptospirosis occur mainly after flood, making it an occupational hazard for sanitary and agricultural workers, as well as a recreational hazard for humans (5). Some pathogenic leptospira species have also been found to be associated with domesticated animals. For example, serovar canicola (*Leptospira canicola*) has adapted itself to canines; therefore, it has become common in many human communities. Although there has been no report of leptospirosis in canines in Thailand, there is a great potential for the transmission of the disease between humans and dogs kept as household pets, unless one is aware of the disease.

*L. canicola* cells used in our study are motile aerobes that are very thin, flexible and spiral-shaped of about 0.1 μm width and 6-20 μm length. *Leptospira* cells are difficult to observe under a light microscope. They can, however, be observed by dark-field microscopy using wet samples. This allows for the determination of agglutination immunoreactivity to be determined. The leptospiral outer membrane or surface antigens can be detected through its agglutination with a homologous [antiserum]. The optimal conditions for its growth and as well, its biology are well documented in the literature (1, 2). Moist environments with a neutral pH are suitable conditions for the survival of leptospira outside the host. The optimal cultivation temperature is approximately 20-32°C. In general, *Leptospira* species are highly susceptible to adverse environmental conditions such as exposure to dry air, chemicals such as chlorine or iodine in detergents, unfavorable pH ( > 8.0 or < 6.5), strong electromagnetic fields and high temperatures (above 40°C).

Magnetic fields (MFs) also affect various biological functions of living organisms, for example, DNA synthesis and transcription (6), as well as ion transportation through cell membranes (7). Almost all living organisms are exposed to magnetic fields from various sources. The geomagnetic field on the surface of the earth is approximately 0.50-0.75 gauss in strength. There have been several studies on the effects of exposure to MFs and several of these have given rise to controversies over the past decades. The growth rate of the Burgundy wine yeast has been shown to decrease when an extremely low magnetic flux density (MFD) of 4 gauss is applied (8). The growth of *Trichomonas vaginalis* is accelerated when it is exposed to 460-1200 gauss (9). The growth rate of *Bacillus subtilis* increases when exposed to 150 gauss and decreases when exposed to more than 300 gauss (10). Similar results were reported for *Chlorella*; an exposure of less than 400 gauss increases the growth, while exposure to 580 gauss decreases the growth rate (11). Several studies point to the MF as a factor influencing the growth and survival of living organisms, which vary at different MFDs (12, 13, 14, 15). Other researchers have studied the effects of MFs on bacteria at the enzyme (16) or genetic (17) level.

To study the efficacy of using magnetic field to control or prevent the growth of *Leptospira*, we applied MF on selected *Leptospira* cells at various intensities and exposure duration levels. We then determined the agglutinating activity of experimental bacteria using dark field microscopy.

## MATERIALS AND METHODS

Pathogenic *Leptospira interrogans*, serovar *canicola* was used in this study. Bacterial cells were grown in the Ellinghausen and McCullough modified by Johnson and Harris [EMJH] liquid medium (2). The bacterial cells were grown at a temperature of $27 \pm 1°C$ in the dark.

A cylindrical permanent ferrite magnet 5 cm in diameter was placed beside 15 ml culture glass tube (less than 1 ml apart) containing 1 ml of a suspension of newly subcultured leptospira cells in the EMJH liquid medium. MF and homogeneity of $140 \pm 5$ mT (northpole) were checked using a teslameter (Hall effect Teslameter digital, order no. 13610.93; Phywe Systeme Gottingen, Germany). The intensity of static magnetic field used in our experiments was chosen on the basis of Genkov et al. (9) findings. Genkov et al. had used more or less this intensity of a constant MF to induce the growth and development of *Trichomanas vaginalis*. For this type of exposure, no shielding against the natural variations of terrestrial MF was required, the value of approximately 0.050 mT is negligible with respect to the MF intensities applied. An experiment using cells not exposed to MF was simultaneously performed as the control, which was placed at a distance of about 100 cm from the exposed group.

In the absence of magnets, MFD was 0.05±0.01 mT. All bacterial samples were exposed to MF for different durations, that is, 0 (control sample), 1, 2, 3, 4, 5, and 6 d. After MF exposure, individual samples were further incubated for 7 d. Immediately after 7 days of incubation, dark-field micrographs were taken using a CCD camera to observe cell development. The growth and agglutination properties using the microscopy agglutination test (MAT) with a homologous antiserum and immunoreactivity were scored as follows:

    4+ = 100 % absence of *Leptospira* cells from the field

    3+ =  75 % absence of *Leptospira* cells from the field

    2+ =  50% absence of *Leptospira* cells from the field

    1+ =  25% absence of *Leptospira* cells from the field

stet

MAT has been commonly used as a diagnostic tool for leptospirosis. This may not be the most reliable test. It, however, is arguably the most appropriate test for this study. The same set of conditions and specimens were used in the experiments, which were repeated twice.

**Atomic Force Microscopy (AFM) and sample preparation** Scanning probe microscopy (SPM) (Digital Instruments Veeco Metrology Group, New York, USA) was used for AFM surface morphology imaging. Images were acquired in the contact mode showing height contours that highlight the spiral shape and fine surface morphology of *Leptospira* cells. An AFM scanner with hardware correction for the nonlinearities of the piezoelectric element was used. The scanner has a maximum xy range of 125 by 125 µm and a z range of 6µm. The cantilevers of $Si_3N_4$, 125 µm long and 35 µm wide with a spring constant of 0.58 $Nm^{-1}$ were used. To locate the area of interest in the samples and identify any bacteria, we used a built-in long-range on-axis microscope, capable of a 5:1 zoom and x 3,500 magnification. Imaging was carried out at scan speeds between 1 and 50 µm/s. Images were acquired at 256x256 pixels. A typical imaging session began using a built-in optical microscope and by moving the x-y table to search for bacterial cells. The AFM cantilever was then moved forward to the surface close to the chosen bacterial cell.

Each sample was prepared using the method described above. It was then dropped on a microscope glass slide and dried in air.

## RESULTS

Figure 1 shows the AFM picture of a *L. interrogans serovar canicola* cell taken with a Digital Instrument Nanoscope IIIa (Digital Instruments Veeco Metrology Group, New York, USA) in the contact mode. The image shows a normal morphology of *L. interrogans serovar canicola*, that is, the spiral shape. It is worth noting that AFM usually reveals the actual roughness of the surface of the bacterial envelope. Other types of microscopy frequently show the surface to be relatively smooth. This technique was also used to observe the surface morphology of bacterial cells before and after the exposure to MF. It should be noted that this image does not demonstrate the rough envelope very clearly. However, it does show the normal bacterial morphology.

## Figure 1

Figure 2 shows some representative dark field micrographs of *L. interrogans serovar canicola* taken at the logarithmic growth phase (at 1:10 dilution of culture samples) and for different durations of MF exposure, that is, 0, 2, 3, and 6 d. After 7 d of incubation, the samples were observed under a dark field microscope and images were taken using a CCD camera. Even though there are some noises in the images, the inhibition of cell growth could be observed. The implications of these observations are significant given the results of other studies(6-17). From Figs. 2A to 2D, one can clearly observe that cell density decreased with exposure time, particularly after more than 3 d. This indicates the decrease in growth rate resulting in the decrease in the number of bacterial cells. This is one of the factors that explain the lower agglutination immunoreactivity, which is there were fewer remaining living bacterial cells to agglutinate.

## Figure 2

Figure 3 shows the dark field micrographs of agglutinated bacterial cells after reacting with the specific antiserum; Fig.3A shows a complete agglutination (100% immuno) and Fig.3B shows 50% agglutination (with only one half of free-living bacterial cells present).

## Figure 3

On the basis of the criteria mentioned at the end of the previous section, the agglutination reactivities of the *L. interrogans serovar canicola* exposed to different intensities of MF are listed in Table 1 (with longer exposure time, the leptospiral bacterial cells demonstrated a lower agglutination immunoreactivity than that of the reference antiserum tested. The end point of reactivity was 50% agglutination (2+)). The agglutination immunoreactivity score decreased with exposure time of *Lectospire cells* as shown in Fig 4. Comparing the MAT results of control *Leptospire cells* (0 d exposure) and those of bacterial cells after exposed to MF, we found that the latter groups (particularly those with longer exposure) showed lower agglutination reactivies. These findings may indicate the presence of a lower amount of agglutinin or number (density) of *Leptospires cells* in the exposed samples than in the control samples. It should be emphasized that the same set of conditions and specimens were used in the experiments that were repeated twice, and the experiments yields exactly the same (semiquantitative) results. The scoring data therefore did not show an error. Once again, each experimental setup, it has one control (nonexposed) group and six exposed groups with different durations of exposure.

## Table1

## Figure 4

Besides the decrease in the number of *Leptospira cells* as the cause of the decrease in agglutination immunoreactivity as mentioned above, the "denaturing effect" of the antigen-antibody reaction may be an other contributing factor to this phenomenon, which can be explained as follows: Typically, antibodies are large soluble protein molecules known as immunoglobins and are produced by B-cells. They bind to specific antigens in a lock-and-key fashion (lock = antibody; key = antigen) (18). Their shape should, therefore, be specific to particular antigens. When a specific antibody encounters an antigen, it will form an antigen-antibody complex through some noncovalent forces such as electrostatic force, hydrogen bond, van der Waal force or hydrophobic force. When a change in what of a single atom occurs, the complex can become unbound. This specificity could be the underlying factor for the denaturation of the antigen-antibody reaction. Under the conditions used in the study, the motion or transfer of any electrons or ions onto the cell membrane could induce an electric current. This current may perturb the other charge particle motion in the cell thus resulting in the loss of binding (19).

# Figure 5

Surprisingly, we observed that some *Leptospira cells* exposed for 3 or more days were longer than the control bacterial cells (see Fig. 5). This preliminary finding probably indicates that there is some disturbance in cell division. More experiments must be carried out to examine and determine the exact mechanism underlying these observed phenomena. Our present explanation for this abnormality in cell division is based on the following: Like most bacteria and archaea, *Leptospira cells* divide symmetrically possibly via the formation of a septum in the middle of the cell (we consider that binary fission is less likely). For the time being, we use AFM in the investigation of division-related morphologies. Recent evidence indicates that synthesized proteins dedicated to cell division are assembled between segregated chromosomes at an appropriate time (20). The key to this assembly is the filamentous temperature exposure sensitive (Ftsz structural) analogue of tubulin (21). DNA damage caused by MF exposure induces mutation, resulting in the abnormal synthesis of *FtsZ*, which in turn could interfere or stop cell division. Similar to previous studies of *Escherichia coli*, *FtsZ* appears to induce the earliest (known) step in cell division. E.coli cells with a mutation of ftsz caused by exposure to certain conditions do not divide. This result in the formation of long filamentous cells that can replicate and segregate their chromosomes (22).

Our finding is at least the first step toward a grater understanding of this the development of diagnostics, treatment, and prevention schemes for bacterium and leptospirosis. We hope that further studies of leptospirosis will lead to this disease in the near future.

## ACKNOWLEDGMENTS

This research was supported in part by the Thailand Research Fund, TRG4580090 and RTA4580005 and MTEC Young Research Group funding MT-NS-45-POL-14-06-G. The support of the Royal Golden Jubilee Ph.D. Program (PHD/0240/2545) to Jirasak Wong-ekkabut and I-Ming Tang is also acknowledged.

TRIAMPO ET AL.					13

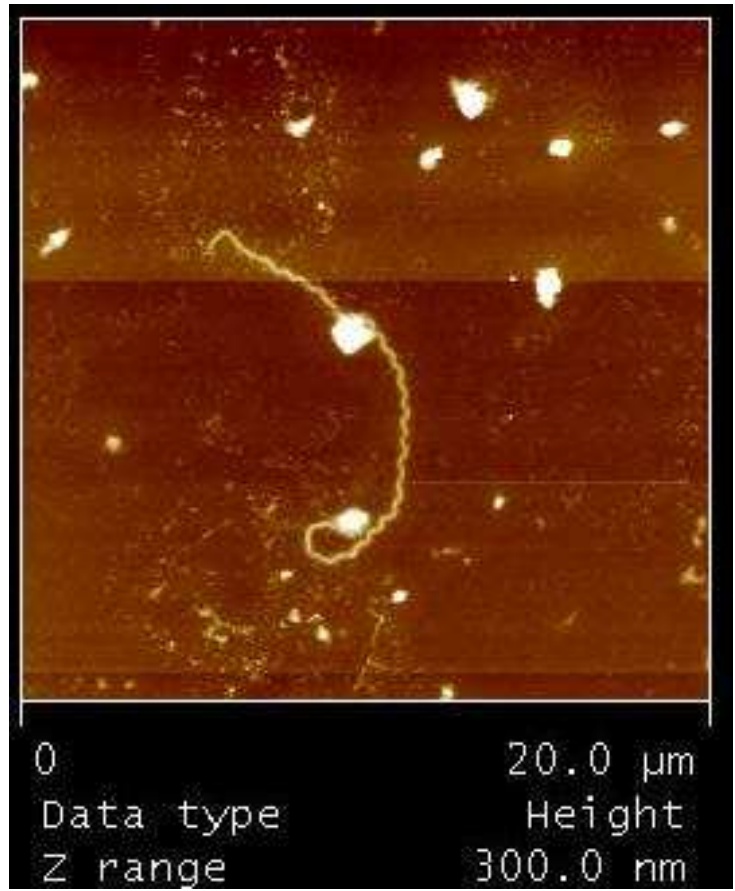

FIG. 1. Atomic force micrograph (AFM) of *Leptospira interogans* serovar *canicola* taken using Digital Instrument NanoScope IIIa in the contact mode under control conditions, that is, without MF exposure. Scan size was 20 µm and scan rate was 1 Hz. It shows a spiral-shaped leptospire of approximately 10 - 20µm.

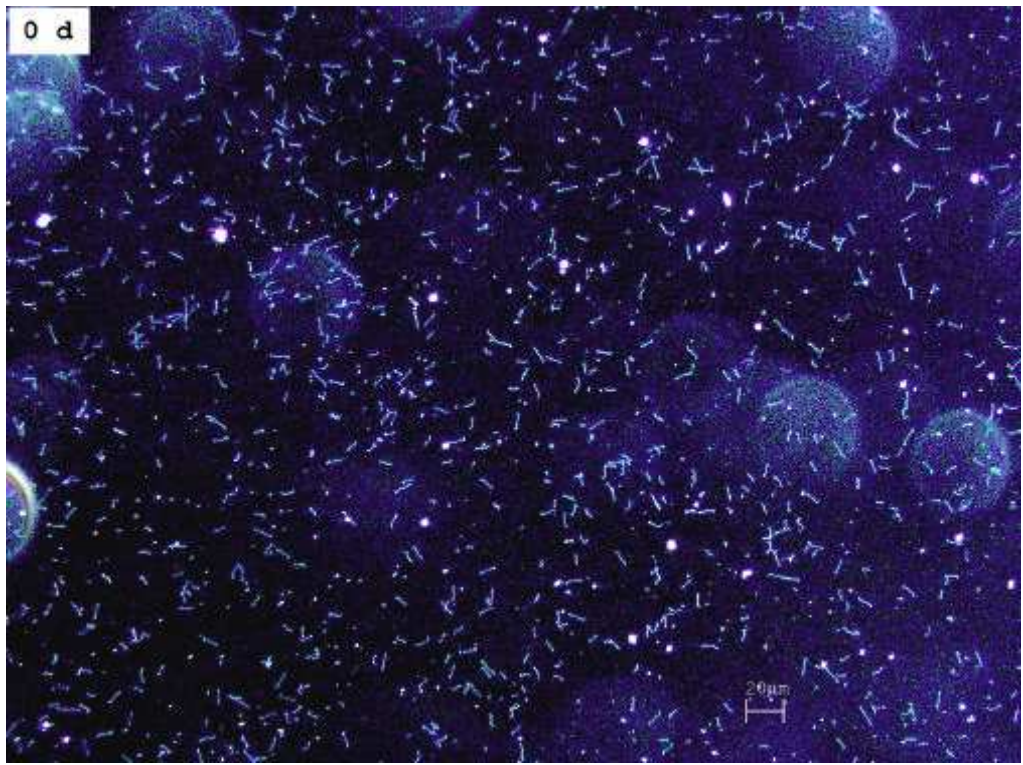

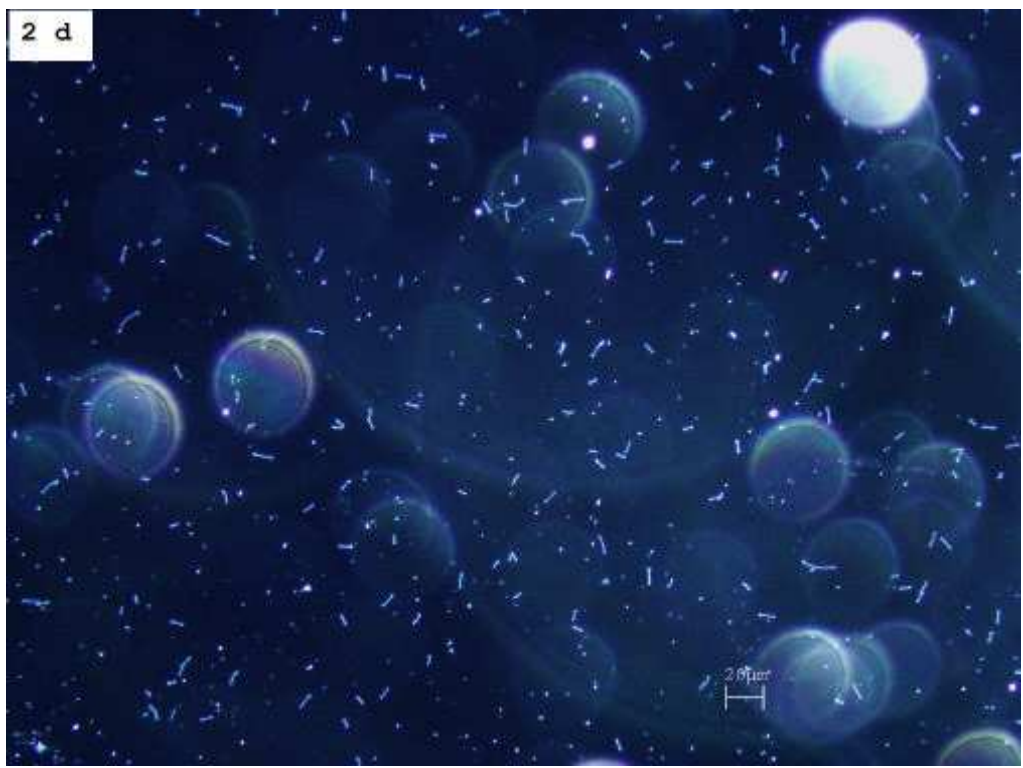

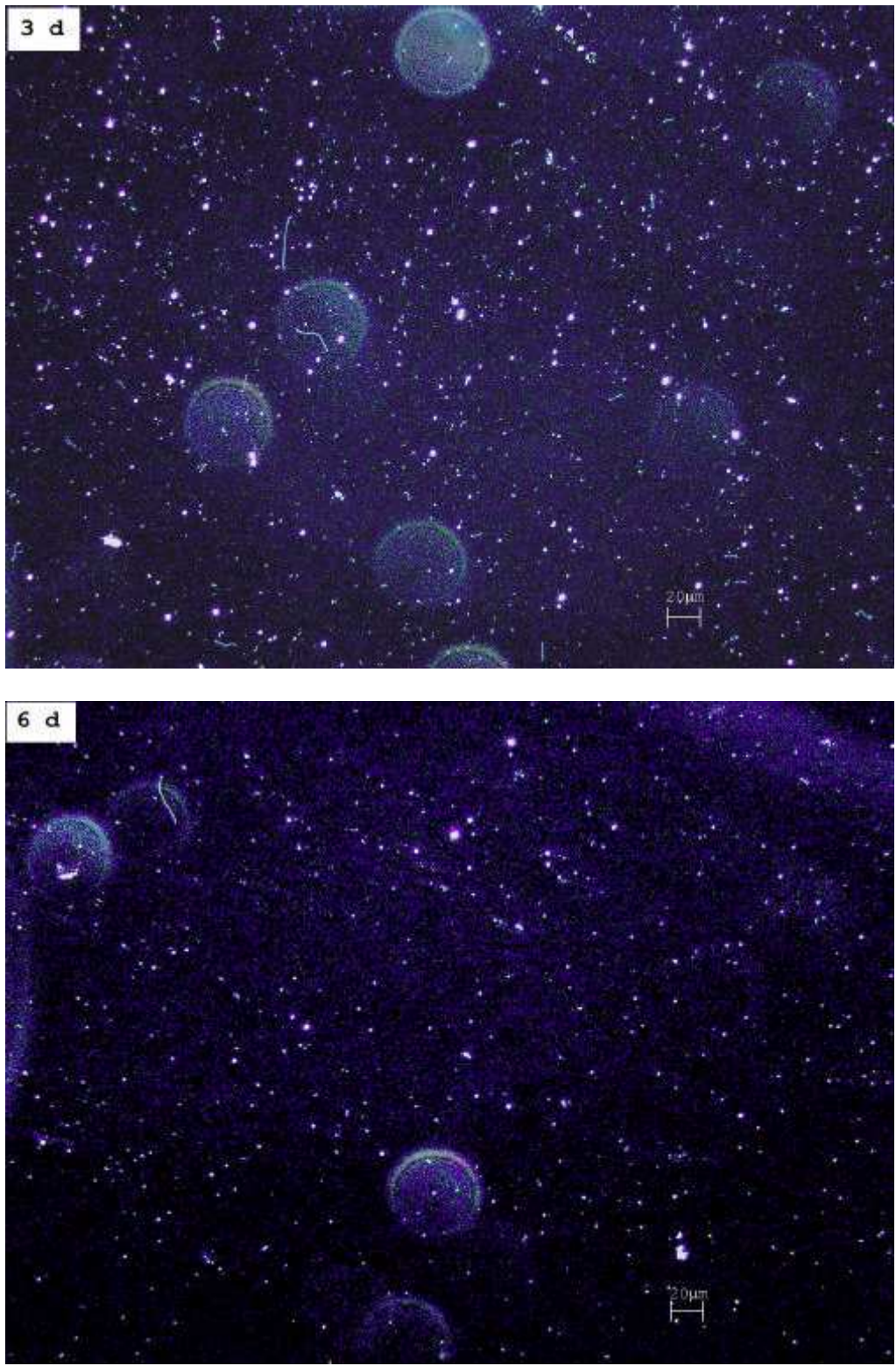

FIG. 2. Dark field micrographs of *Leptospira interrogans* serovar *canicola* exposed to MF for different durations. The images were taken at the log phase of each experimental culture sample (diluted 1:10 of original).

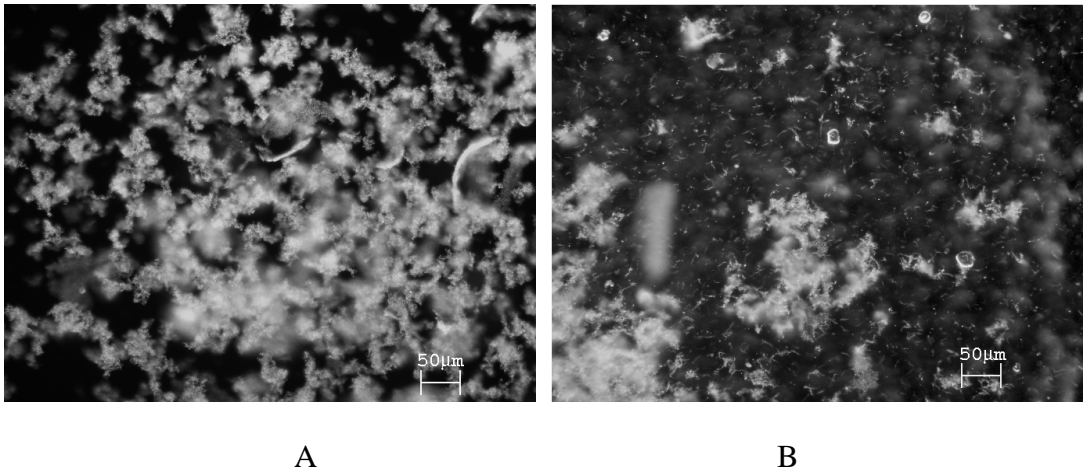

A  B

FIG. 3. Dark field micrographs of agglutinated bacterial cells after reacting with homologous antiserum, showing complete agglutination (100% reactivity; A) and 50% agglutination with one-half of free-living bacterial cells remaining (B).

TABLE 1. Agglutination characteristics of leptospires after magnetic field exposure for various durations.

| Exposure duration (d) | 1:50 dilution | 1:100 dilution | 1:200 dilution | 1:400 dilution | 1:800 dilution | 1:1600 dilution | 1:3200 dilution |
|---|---|---|---|---|---|---|---|
| 0[a] | 4+ | 3+ | 2+ | 2+ | 2+ | 2+ | 1+ |
| 1 | 3+ | 2+ | 1+ | - | - | - | - |
| 2 | 3+ | 2+ | 1+ | - | - | - | - |
| 3 | 2+ | - | - | - | - | - | - |
| 4 | 2+ | - | - | - | - | - | - |
| 5 | 1+ | - | - | - | - | - | - |
| 6 | NA | - | - | - | - | - | - |

a   Representative sample of control unexposed leptospires showing a higher MAT titer (1:1600) than exposed samples for various durations.

NA indicates no agglutination occurred.

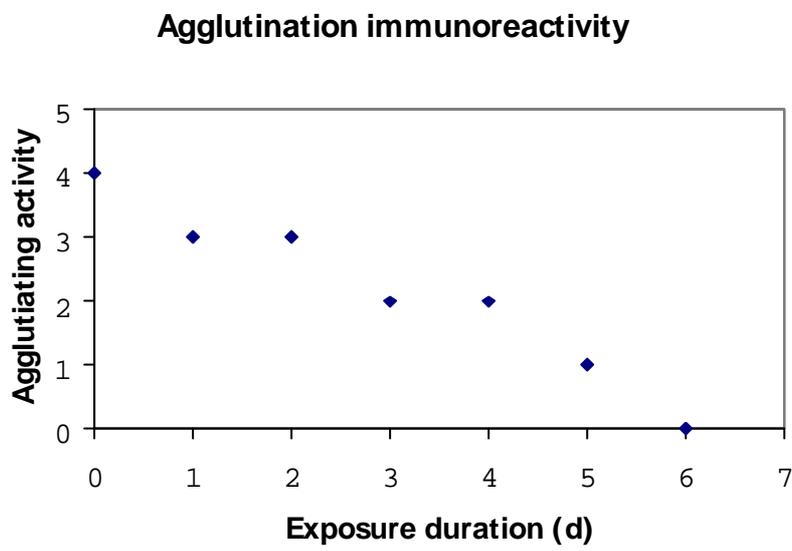

FIG. 4. Plots of data shown in Table 1.

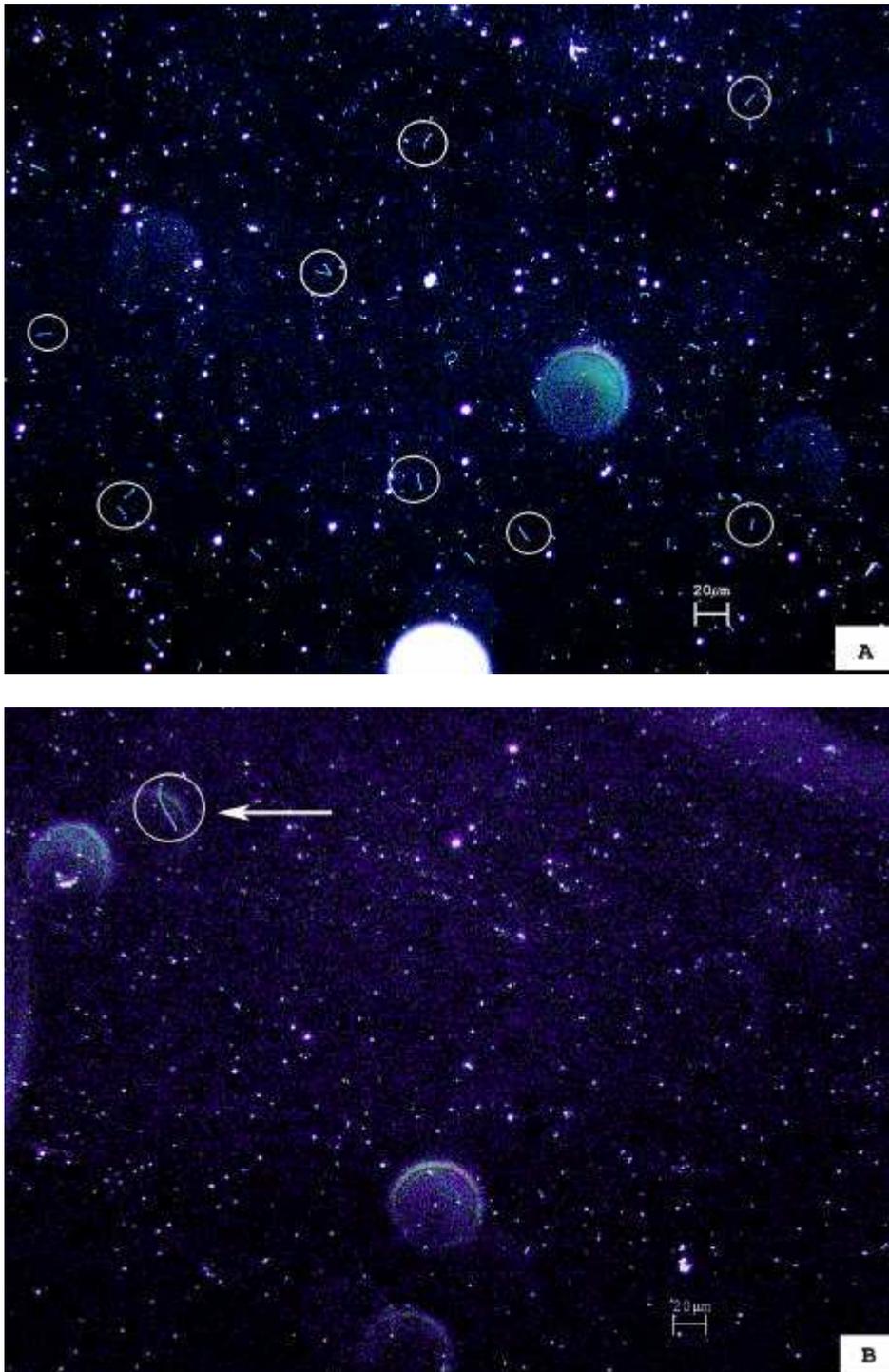

FIG. 5. Dark-field micrographs of *L. interrogans serovar canicola* taken at the same magnification (x200). Control sample unexposed to magnetic field; the leptospires have an approximate length of 10-20 μm (A) compared with magnetic field-exposed leptospires (B) with some cells longer than others. Circles indicate individual bacterial cells.